%% file: ms.tex
\documentclass[namedreferences]{solarphysics}

\usepackage{mathtools}
\usepackage[hyperref,optionalrh,showbiblabels]
{spr-sola-addons} 
\usepackage{graphicx}
\usepackage{color}   
\usepackage{ulem}
\def\UrlFont{\sf}    

\def\arp#1{{\color{cyan}{#1}}}  

\begin{document}

\input defs

\begin{article}
\begin{opening}

\title{Efficient and Automated Inversions of Magnetically-Sensitive Forbidden Coronal Lines: CLEDB - The Coronal Line Emission DataBase Magnetic Field Inversion Algorithm}

\author[addressref={aff2,aff1},corref,email={arparaschiv@ucar.edu}]{\inits{A.R.}\fnm{Alin~Razvan}~\lnm{Paraschiv}}

\author[addressref=aff1]{\inits{P.G.}\fnm{Philip~Gordon}~\lnm{Judge}}

\address[id=aff2]{National Solar Observatory, 3665 Discovery Dr, Boulder, CO 80303,  USA}

\address[id=aff1]{High Altitude Observatory, National Center for Atmospheric Research, PO Box 3000, Boulder CO 80307, USA}

\begin{abstract}
We present CLEDB, a ``single point inversion'' algorithm for 
inferring magnetic parameters using $I,Q,U,$ and $V$ Stokes 
parameters of forbidden magnetic dipole lines formed in the solar corona.
We select lines of interest and construct databases of Stokes parameters 
for combinations of plasma thermal and magnetic configurations. The size and complexity of such 
databases are drastically reduced by taking advantage of symmetries. 
Using wavelength-integrated line profiles, each of which might be decomposed beforehand into several line-of-sight components, we search for nearest matches 
to observed Stokes parameters computed for the elongation 
corresponding to the observed region. 
The method is intended to be applied to two or more lines 
observed simultaneously. The solutions initially yield magnetic orientation, thermal 
properties, and the  spatial position of the emitting plasma in three dimensions.
Multiple possible solutions for each observation are returned, including 
irreducible degeneracies, where usually sets of two solutions are compatible with the two input $I,Q,U,$ and $V$  measurements. In solving for the scattering geometry, 
this method avoids an additional degeneracy pointed out by \citet{Dima+2020}.
The magnetic field strength is separately derived from the simple ratio of observed to database Stokes $V$ data, after the thermal properties
and scattering geometry solutions have been determined. 
\end{abstract}

\end{opening}

\section{Introduction}
     \label{S-Introduction} 

Our need to measure  the magnetic field
threading the solar corona has never been more urgent. Society
depends on electrical 
infrastructure in space and on the ground to an unprecedented and ever-increasing degree. The greatest single source of electrical perturbations on the 
Earth is the Sun, as has been known for at least a century and a half.   Variable high energy radiation, ejection of 
magnetized plasma, the interactions of streams in adjacent sectors of the solar wind, all these lead to potentially dangerous effects 
on a technologically-dependent society. These 
and other issues are discussed in a variety of 
 monographs, white papers and reviews
\citep[e.g.][]{Billings1966,
2001STIN...0227999J,Eddy2009,
2013SoPh..288..467J, 2017SSRv..210..145C,Ji+Karpen+others2020}.

The urgency of finding a reliable 
method to measure coronal magnetic fields
arises from the fortunate 
conjunction of 
three unique observational 
opportunities. The Daniel K. Inouye Solar Telescope (DKIST, formerly ATST,
see \citealp{Rimmele+others2003,rimmele+others2020}),
the Parker Solar Probe (PSP, formerly Solar Probe Plus, see 
\citealp{PSP}) and the Solar Orbiter mission (SolO,  \citealp{Orbiter,Orbiter2})  are all now 
operational.  
While the PSP and SolO orbit the Sun beyond 9 solar radii (R$_\odot$), sampling \textit{in-situ} plasmas, neutral particles and magnetic fields, 
the 4 meter DKIST observatory will be able to measure 
components of magnetic fields at elongations
$y \lesssim 1.5$ R$_\odot$.  Such a large aperture, coronagraphic  telescope 
operating from the peak of Haleakala marks 
a huge step up from  early feasibility 
efforts with far smaller telescopes 
\citep[e.g. Evans Solar Facility, SOLARC,  COMP;][]{Lin+Penn+Tomczyk2000,Lin+Kuhn+Coulter2004,CoMP}. We also eagerly anticipate synoptic measurements with UCOMP \citep{Tomczyk+landi2019} and the 1.5 meter 
aperture coronagraph (\href{https://www2.hao.ucar.edu/cosmo/large-coronagraph}{www2.hao.ucar.edu/cosmo/large-coronagraph}) of the COSMO suite of instruments, currently under review by the community.

In the current work, we describe
a numerical method for 
studying  magnetic signatures imprinted in the polarized light from magnetic dipole (M1) 
lines emitted at visible and infrared wavelengths by the corona.  These M1 lines are formed in the saturated Hanle effect regime and are optically thin across the corona. Concerns have been
expressed regarding line-of-sight (LOS) confusion 
\citep[e.g.][]{2013SoPh..288..467J,2020SoPh..295...98S}. Mathematically, null spaces exist where variations in 
vector magnetic fields have no effect on the emergent spectra. Contributions to the familiar observed Stokes parameters $I,Q,U,$ and $V$  come from 
different regions along the LOS.

\figsym

Single-point  algorithms, like the one presented here must be considered a first step  until stereoscopic  observations, involving spacecraft in orbits significantly away from the
Earth-Sun line
of the corona, become available. Alternatively  
the Sun's rotation might be used to 
try to probe the 3D coronal structure using stereoscopy
\citep[e.g.][]{Kramar+2016}, assuming rigid coronal rotation  over periods of days or longer.  The corona may or may not comply with this
assumption, and  stationary structures which do comply may be of
limited physical interest anyway.

Thus, our purpose is
to present a method along with a python-based tool to allow coronal observers with DKIST and other telescope systems, to obtain 
a first estimate of properties of
the emitting plasma including components of the vector magnetic field. 
Our primary 
simplification is 
\begin{quote}
\textit{to seek solutions 
for the emitting plasma assuming it is dominated 
by one location along the line-of-sight.}
\end{quote}
For practical purposes, 
such ``locations'' make sense if they span LOS lengths 
 smaller than, say, 0.1$R_\odot$.   Naturally, without  observations from 
 a very different LOS 
 in the solar system, the 
 measurements represent differently-weighted averages of physical conditions along each LOS.
 These solutions inherently possess well-known ambiguities arising from specific symmetries 
 associated with the 
 line formation problem, as shown for example in Figure~\ref{fig:sym}.  Therefore, our inversion scheme allows one to identify, but not necessarily resolve, all 
ambiguities from a set of observed Stokes profiles, as revealed in Section~\ref{S-CLEDB}. Section~\ref{S-discussion} provides a summary of the findings, and discusses multiple emission
locations as suggested by multiple components in the emission lines, where solutions for each can, in principle, be obtained.

\section{Review of the Formation of Forbidden Coronal Lines}\label{S-formalism}
\subsection{Emission Coefficients in Statistical Equilibrium}

We adopt the formalism and notation of \citet{Casini+Judge1999}, that expands upon
earlier work on loosely related topics 
\citep[e.g.][]{Sahal,1991A&A...244..391L,landi82}. We must solve 
for the magnetic substate
populations of the radiating ions assuming statistical equilibrium (SE). 
The problem is cast into
the framework of 
spherical tensors to take advantage of  geometrical symmetries \citep[see Chapter 3 in][]{Landi}. Magnetic Dipole (M1) coronal lines form under regimes 
where Zeeman frequency splittings are of order of the classical Larmor frequency $\nu_L$, $h\nu_L =\mu_0\cdot B$ and are far smaller than the Doppler widths $\Delta\nu_D$, and where 
the Einstein - A coefficients $A_{JJ_0}$ are, in turn $ \ll \nu_L$.  The first inequality 
permits an accurate Taylor expansion of line profiles in terms of
the small quantity $\delta
= \nu_L/\Delta\nu_D \ll 1$
\citep{Casini+Judge1999}.
The second defines the ``strong-field limit of the Hanle effect'' in which coherences between 
magnetic sub-states of the decaying level are negligible in the magnetic-field reference frame. 
From the solutions
to the SE equations, the emission coefficients for the Stokes 
vector $(S_0,S_1,S_2,S_3)^T$ $\equiv 
(I,Q,U,V)^T$ are then  (see Equations~35a-35c in
\citet{Casini+Judge1999}):
\begin{eqnarray}
\varepsilon_0^{(0)}(\freq,\hat {\bf k})
&=&\epsilon_{ J J_0}\,\phi(\freq_0-\freq)
	\left[1+D_{J J_0}\,\sigma^2_0(  J)\,
	{\cal T}^2_0(0,\hat{\bf k})\right]\;,
\label{eqn:eps00} \\
\noalign{\smallskip}
\varepsilon_i^{(0)}(\freq,\hat {\bf k})
&=&\epsilon_{ J J_0}\,\phi(\freq_0-\freq)\,D_{J J_0}\,
	\sigma^2_0(  J)\,
	{\cal T}^2_0(i,\hat {\bf k})\;,
	\qquad (i=1,2)
\label{eqn:epsi0} \\
\noalign{\smallskip}
\varepsilon_3^{(1)}(\freq,\hat {\bf k})
&=&-{\textstyle\sqrt{\frac{2}{3}}}\,\freq_{\rm L}\,
	\epsilon_{ J J_0}\,\phi'(\freq_0-\freq)
	\left[\bar{g}_{  J,  J_0}+E_{J J_0}\,
	\sigma^2_0(  J)\right]
	{\cal T}^1_0(3,\hat {\bf k})\;,
\label{eqn:eps31} 
\end{eqnarray}
\noindent  where 
the M1 transition occurs from level  with angular momentum 
$J$ to  $J_0$, and where 
\begin{equation}
   \epsilon_{ J J_0} ={\frac{h \freq}{4\pi}}\,N_{ J}\, A_{JJ_0}.
\end{equation}	 
$\phi(\freq_0-\freq)$ is the (field-free) line
profile (in units of Hz$^{-1}$),
with 
$\int_0^\infty\phi(\freq_0-\freq) d\nu =1$
and $\phi'(\freq_0-\freq)$ denotes its first derivative with respect to $\nu$.  

When integrated along a specific LOS, the 
expressions for the emission coefficient 
$\varepsilon^{(i)}_i(\nu,\hat {\bf k})$, with units of
erg~cm$^{-3}$ sr$^{-1}$s$^{-1}$, 
yield the emergent Stokes vectors from the 
corona. $\epsilon_{ J J_0}$ is the usual coefficient 
for the frequency-integrated isotropic 
emission only from the line, ignoring 
stimulated emission, where the $D_{ J J_0}$ and $E_{ J J_0}$ coefficients are dimensionless parameters associated with the polarizability of the two atomic levels.

The superscripts on $\varepsilon^{i}$ are the leading orders in the Taylor expansion of the line profile
\begin{equation}
\phi(x+dx) = \phi(x) + \sum_{j=1,...} \frac{d^j}{dx^j} \phi(x) \cdot dx^j \ldots
\end{equation}
with $dx \propto \delta=\nu_L/\Delta\nu_D \ll1$. Second order terms 
in $\delta$ are negligible for 
weak coronal fields and broad
line profiles.  Lastly, here it is assumed 
that 
the photospheric radiation
is spectrally flat across the corona line profiles \citep{Casini+Judge1999}. 

\subsection{Physical Interpretation}

These equations are readily
understood physically.
The leading order in the $IQU$ Stokes signals is zero, for Stokes $V$ it is one.  $IQ$ and $U$ arise from a combination of thermal emission and scattering of photospheric radiation, both include the populations $N_J$ 
and the atomic alignment 
$\sigma^2_0(J)$. Both 
quantify local solutions
to the SE equations, entirely
equivalent to solving for
populations of magnetic sub-states \citep{House1977,Sahal}. 
Alignment 
is generated entirely by the anisotropic irradiation of ions by the underlying solar photospheric radiation.   Information on the magnetic field in $IQU$ is contained 
implicitly in $\sigma^2_0(J)$ and is independent of
magnetic field strength, as corresponding to the mathematical statement of the 
strong field limit. In contrast, Stokes $V$ for the M1 lines is 
formed entirely through the 
Zeeman effect, modified by
the alignment factor \citep{Casini+Judge1999}. 
When the atomic alignment 
factor is zero, the expression for Stokes V
reduces to the well-known 
``magnetograph formula''
of the Zeeman effect, to 
first order.  The leading 
terms for $QU$ in the 
Zeeman effect are only second order in $\delta$ leading them to be considered negligible in coronal cases due to small B magnitudes. Together with the first order term in $V$ they form the basis of most solar ``vector polarimeters'' \citep[e.g.][]{Lites2000}.

The coefficients $D_{JJ_0}$,
$E_{JJ_0}$ 
are properties fixed by quantum numbers $J$ and $J_0$
of the two atomic levels. f
$D_{JJ_0}$ is fixed by 
$J$ and $J_0$, but $E_{JJ_0}$ 
also depends on each level's ``Land\'e g-factor'' that are used to build 
``effective Land\'e factors'' of the transition, $\bar{g}_{J,J_0}$.   
The Land\'e g-factors also depends on 
quantum numbers other than 
$J,J_0$, like
the mixing of atomic states 
and orbital and spin 
angular momenta. These can be measured or 
may be computed using an atomic structure calculation. See, for example, new calculations of special relevance to this work by \citet{Schiffman+others2020}.

 The Taylor expansion of $\varepsilon_i^{(o)}(\nu,\uvec{k})$ with frequency has leading orders $o=1$ when $i=3$ and $o=0$ otherwise \citep{Casini+Judge1999}.

The terms $N_J$ and $\sigma^2_0(J)$, 
the population and alignment of level with total angular momentum $J$,  
are solutions to SEs.  These solutions, which are linear combinations of the populations of magnetic substates of level $J$ \citep[e.g.,][]{Sahal}, 
depend both on the 
scattering geometry, 
the magnetic unit vector 
$\uvec{b}$, and the plasma 
temperature and density.
The atomic alignment $\sigma^2_0(J)$
is created by the bright, anisotropic 
photospheric cone of radiation seen by the coronal ions, and destroyed by collisions with plasma particles 
having  
isotropic distributions.  
This ``atomic polarization'' is 
modified by the magnetic field as
the ion's magnetic moment precesses around the local B-field.  
The appearance of 
$\sigma^2_0(J)$ in the SE 
equations underlying 
Equations~\ref{eqn:eps00}-\ref{eqn:eps31} show that 
linearly polarized light in the corona originates  
from atomic polarization, and also that the intensity and Zeeman-induced 
circular polarization are modified by it. 

\subsection{Scattering Geometry}

Finally, the spherical tensors ${\cal T}^K_0(i,\hat {\bf k})$  
define the geometry of the scattering 
of solar radiation
for Stokes component $i$ from the coronal  plasma.  The tensors play
no role in the SE calculations, as is
readily appreciated, in that the SE states cannot depend on the observer.  Figure~\ref{fig:sym}(b)
shows instead how the solutions 
depend on $\vartheta_B$, the angle between the local magnetic field and
the radius vector $\uvec{r}$ to the local vertical of the Sun ({\it l.v.s}).

\section{CLEDB, a Database Approach for ``Single-Point Inversions''}\label{S-CLEDB}
\subsection{The CLEDB Algorithm}

\figcledb

The Coronal Line Emission DataBase (CLEDB) inversion algorithm is created to harness all available information in polarization measurements of the corona to infer local plasma properties and vector magnetic fields. A non-commercial open-source python-based code package of CLEDB, designed for both personal computer jobs and SLURM (Simple Linux Utility for Resource Management) enabled research computing jobs, is freely available online. More information about the code, package, and method documentation along with persistent links are found in the data availability section.
The algorithm uses the equations and framework described in Section \ref{S-formalism} together with symmetries and line profile properties to extract magnetic and thermal information from measured Stokes parameters through a search of a database of computed Stokes parameters.  

A single emission line does not contain sufficient information for a full inversion. This will become clear below, but see 
\citet{Plowman_2014}, \citet{Dima+2020}, and \citet{Judge+Casini+Paraschiv2021} for detailed discussions. Therefore, the CLEDB approach is primarily designed for two or more coronal lines. A secondary code branch will be used to derive basic thermal parameters and LOS magnetic fields only, when Stokes observations of 1-line are provided instead of 2-line using analytical approximations incrementally developed by \citet{Casini+Judge1999}, \citet{Plowman_2014} and \citet{Dima+2020}. 

In the CLEDB 2-line configuration, solutions that are deemed a good fit, currently by using a reduced $\chi^2$ metric, are returned along with database model magnetic, geometric and thermal parameters as acceptable solutions to the inverse problem.

The algorithm seeks thermal and magnetic conditions 
from a single point along the LOS.   This is
a gross oversimplification 
in general, but it is well known that coronal images 
frequently reveal discrete structures, such as in polar plumes or more especially in  loops over magnetically active regions.  These are the regions of great interest
for space weather disturbances at the Earth \citep{Ji+Karpen+others2020}. 
However, in cases such as the quiet Sun, the emission is distributed diffusely and our method will represent some 
poorly-defined average 
of quantities along the LOS.  We explore this
assumption below.
In essence, we replace the integrals of equations \ref{eqn:eps00}-\ref{eqn:eps31}
   over the LOS
with a 1-point quadrature using a length scale $\ell$. For convenience, we choose $\ell=1$ and use henceforth,
\begin{equation} \label{eq:S}
     S_i(\freq, \uvec{k}) \equiv \epsilon_i^{(o)} (\freq, \uvec{k}),
\end{equation}
which is the emergent Stokes parameter of the emission line for a path length of 1 cm along $\uvec{k}$.   

Even with this simplification,
there is always some ambiguity in the solutions owing to inherent 
symmetries.  Our algorithm 
therefore returns all such 
solutions deemed to be compatible with the
data. 

\tabangles

\subsection{Frames of Reference}

Figures~\ref{fig:sym},~\ref{fig:spheresm}
and Table~\ref{tab:angles}
define various angles 
in terms of a Cartesian
reference frame with its origin at the center of the Sun.  
The axes 
 $\mathbf{ \hat x},
\mathbf{ \hat y},
\mathbf{ \hat z}$
point along the Sun center-observer line, the E-W direction and S-N
direction relative to the Sun's 
rotational axis in the plane-of-the-sky.   We adopt the reference direction for linear polarization 
to be along the $z$- axis (vertical). 
This corresponds to the direction of a linear polarizer measuring $\frac{1}{2} (I+Q)$ \citep[see p. 19 of][]{Landi}.

Two unit vectors $\uvec{r}$ and $\uvec{b}$ specify the direction of 
the center of the cone of photospheric 
radiation and magnetic field, and a third $\uvec{k}$ specifies the LOS\footnote{Strictly speaking, the $\uvec{k}$ vectors drawn at points $O$ and $P$ 
are not quite parallel, but here we ignore this small difference, as they 
are $\lesssim0.5^\circ$ 
different when observing plasma with an elongation of $\lesssim 2R_\odot$. See Figure~\ref{fig:spheresm} and Table \ref{tab:discretization}.}.

We define two reference frames, the ``solar" frame and the ``observer'' frame.  All angles in the solar  frame
are specified as Greek lowercase letters. 
Two more angles are defined 
in uppercase, defined relative to the observer.
$\Theta_B$ is the angle
between the LOS vector $\uvec{k}$ and 
$\uvec{b}$.  The angle 
$\Phi_B$ follows from our adoption of a reference direction parallel to the $\uvec{z}$- axis.
With this geometry, 
\begin{equation} \label{eq:Phib}
    \Phi_B = \pi - \gamma_B= + \frac{1}{2} \arctan \frac{U}{Q}
\end{equation}
for each line with measurable $Q$ and $U$.  

\tabsols

\subsection{Symmetries to Minimize Numerical Work}

\figspheresm

Frequency-dependent line profiles are not
required because we know \textit{a priori} that, under the single-point contribution assumption, the 
profiles for $I,Q,U$ are identical, namely the zeroth-order term in the Taylor expansion. The leading order in the $V$ profile is the first order
term $\propto dI/d\nu$. 
Therefore we need only create a database of quantities 
$\varepsilon^{(i)}(\nu,\uvec{k})$ appropriately integrated over frequency,
\begin{equation}\label{eq:si}
S_i=\int_{\mathrm{line}}[ I(i,\lambda)-I(i,\lambda_c) ] \;d\lambda.
\end{equation}
The integration for an observed set of Stokes $O_i$ follows the same formalism, when subtracting $\lambda_c$ continuum emission and setting any Doppler shifts to zero. The integral for $V$
requires weights of opposite sign at either side of the line center. If two or more components are
 identifiable
in the $I(\nu,\uvec{k})$ profiles, for example by 
multiple fits of Gaussian profiles, 
the components can be extracted beforehand, and searches made for each component. 

Even with these simplifications, minimal implementations of a search algorithm would generate
databases of impractically large
sizes. A 3D 
Cartesian grid built around a 
quadrant around the solar disk
(Figure~\ref{fig:spheresm}), would demand computation of 
the $S_i$
parameters at each of, say,   $50\times50\times50$ ``voxels''. Each such voxel requires a grid of magnetic vectors 
$\mathbf{B}=(B_x,B_y,B_z)^T$, the LOS components of
velocity field $v_x$, temperature $T$, density
$n_e$, elemental abundance $\mathcal{A}$, and 
a spectroscopic turbulence representing unresolved non-thermal motions $v_T$.  With 
over $10^5$ voxels, the 
number of database entries 
would exceed 
$N_C \ge 10^{13}$, 
using just 10 values for each of the 
magnetic and thermodynamic
variables listed above.  But 
the database size can be
dramatically reduced based upon the following arguments:
\begin{enumerate}
    
    \item Observations are subject to the geometrical rotation of the $Q$ and $U$ profiles using equations~\ref{eq:qrot} and \ref{eq:urot}. 
All $QU$ data can be rotated around the $x$-axis by the azimuth angle $-\alpha$, as shown in Figure~\ref{fig:sym}(a).
Database searches can then be limited to those LOS within 
the $z=0$ plane instead of the entire 3D volume; e.g. point Q is different from point P in Figure~\ref{fig:sym}(a) only by the $\alpha$-rotation of the Q and U Stokes profiles.
Afterwards, matching magnetic vectors are simply rotated back by $+\alpha$. The $I$ and $V$ Stokes parameters are invariant
to rotations about the LOS ($x$- axis). 
    
    \item We need only to search along the LOS $x$- direction using $n_y$ separately stored database files for each observation with an observed elongation $y$ closest to the computed CLEDB height $y_0$,
    minimizing CPU and memory requirements.
    \item We suggest adopting line pairs from a single ion,
    eliminating the need to account for relative abundances and  differential temperatures along each LOS. However, it is possible to use different ions, even of different elements, although this is not advisable for reasons that will become clear below. (see \citet{Judge+Casini+Paraschiv2021} for detailed discussions.)
    \item We can compute the 
    Stokes parameters and store them 
    for a single field strength
    $B=|{\mathbf B}|$. We then compute the ratio between  
    the computed and the observed values of circular polarization. This simplification results from the strong field limit of the Hanle effect. In other words, CLEDB will solve for the geometry, thermal, and magnetic orientation, and afterwards scale the magnetic field strength using Zeeman diagnostics (equation \ref{eqn:eps31}). 
\end{enumerate}
Thus, in this example, the CLEDB scheme's database will encompass $N_C\approx 10^6$
entries for each of the $n_y$ database computed elongations, as shown in Table~\ref{tab:discretization}. The numbers quoted in this example are not absolute and represent just a starting point. In CLEDB the database parameter configuration is a user editable feature when building databases within the CLEDB\_BUILD module. 

The first simplification is 
equivalent to a rotation of our choice of
reference direction for linear polarization.  The  $Q$ and $U$ parameters fed to the search algorithm are simply
\newcommand\hp{\hphantom{-}}
\begin{eqnarray} \label{eq:qrot}
Q_\alpha&=& Q\cos2\alpha -  U\sin2\alpha,\\
U_\alpha&=&Q\sin2\alpha + U\cos2\alpha. \label{eq:urot}
\end{eqnarray}
The preference for lines belonging to single ions described in Point 3 is not a serious restriction,
because ions of the $np^2$ and $np^4$ iso-electronic sequences with $n=2$ and $n=3$, such as Fe~XIII, possess two M1 lines in the $^3P$ ground terms whose dependencies of $\varepsilon_i^{(o)}(\freq,\hat {\bf k})$  on electron temperature are essentially identical, determined by collisional ionization equilibrium. 
The Fe~XIII 1.0747 $\mu$m  and  1.0798 $\mu$m  line pair has served as the primary target for previous instruments 
\citep{Querfeld1977,CoMP},
and remains a prime candidate for new observations with DKIST.
Point 4 entails the  significant benefit of 
finding solutions 
which depend on higher signal \textit{wavelength-integrated} Stokes profiles (equation \ref{eq:si}), rather than noisier differences, 
    of Stokes $V$ profiles (cf. Equation~11 of \citealp{Dima+2020}). We note that the accuracy of database vs. observation scaling is dependent on LOS effects that are not currently fully quantified, as can be seen in the $\chi^2$ values of Table \ref{tab:table1} that indicate overfitting.

\figflow

The search over angles can then be further restricted. We use the ratio $U_\alpha/Q_\alpha$ to estimate the azimuth angle $\Phi_B$ modulo $\pi/2$ for every M1 line (see Figure~\ref{fig:sym}).  
In the database we adopt grids for the magnetic field vector in spherical coordinates at the point $P$
for angles $\phi$ and $\theta$ in Table~\ref{fig:spheresm}.
For each $\Phi_B$, the $\phi$ and $\theta$ angles are 
related by their definitions by:
\begin{equation} \label{eq:qu}
\tan \Phi_B =\tan\theta \sin \phi.
\end{equation}
Ultimately we are left only to
search a 4-dimensional discretized
hyperspace for each 
elongation $y$, to identify 
matching values of 
$n_e$, $x$, $\phi$, and $\bar\theta(\phi)
$ remembering that B is scaled afterwards, as discussed above.
Here $\bar\theta(\phi)$ includes only
values of $\theta$ compatible with 
equation~\ref{eq:qu}. 

In our Table \ref{tab:table1} example of CLEDB sorting, we simply presort the 10 values of $\bar\theta$ in the numerical grid that are most compatible with Equation~\ref{eq:qu}. We see that solutions are degenerate in pairs of two in terms of supplementary $\Phi_B$ and complementary $\Theta_B$ angles. The number of presorted $\bar\theta$ solutions is configurable via CLEDB controlling parameters. Interpolation is of course possible, but it is not currently implemented due to the yet unknown effects of potential uncertainties.

Yet more computational savings are made noting that the electron densities $n_e$ are strong functions of $r$ because of stratification and solar wind expansion.
Thus, we can reasonably seek solutions of a fixed analytical form for $n_e(r)$ as shown in  Table~\ref{tab:discretization}. The function 
\begin{equation}
n_0(r) = 3\cdot 10^8 \cdot \exp \left(- \frac{r-1}{h}\right) +  10^8\cdot(0.036 r^{-3/2} + 1.55r^{-6}),
\end{equation}
has $r$ in units of $R_\odot$, and  scale height $h=0.0718R_\odot$ ($\equiv$ 50 Mm), where the second term is the formula of Baumbach \citep{Allen1973}. 
The grid-sizes that we have used for testing, and we consider a reasonable starting point are given in Table~\ref{tab:discretization}. The resulting density is given by a smaller array of say 15 discretized values centered on the base $n_o$ electron density, which span orders of magnitude of -2 to 2 in logscale.

\tabdisc

We used the reduced $\chi^2$
metric as a goodness of fit,
with Stokes 
`observations' $S_i$ taken 
from values on the database grid.
Then we write $\chi^2$ as the sum of
\begin{eqnarray} 
    \label{eq:chisq}
    \chi^2_{\text{\tiny IQU}} =& \dfrac{1}{d - p}&
    \left [\sum_{i=0,1,2} { \frac{(S_i - O_i)^2}{\sigma^2_i} }\right ]\text{ and}\\
    \chi^2_{\text{\tiny V}} =& \dfrac{1}{d - p } &\frac{(S_3 -O_3)^2}{\sigma^2_3} \label{eq:chisq2}
\end{eqnarray}
where $O_i$ and $\sigma_i$ are for the observed Stokes $I,Q,U$ parameters, and $O_3$ and $\sigma_3$ correspond to Stokes $V$. Here $\sigma^2$ is a variance associated with noise, not to be confused with
the alignment $\sigma^2_0(J)$ which always is specified by $J$. The distribution of noise in $O_i$ is normal with standard deviation $\sigma_i$. The
rms noise is added to $\sigma_i$ as a function 
of the number of photons
detected in the line. We normalize  
the set of 8 Stokes parameters with respect to the Stokes $I$ parameter corresponding to the strongest line in the set, in order to bypass the need for absolute intensity calibrations.
Here, $d = 4\,n_{line}-1$ is the number of independent data points.
The number of free parameters in the model is $p=4$.
With $d=7$ for two lines, the  factor
$(d-p)^{-1}$ in Equation~\ref{eq:chisq}
is $\frac{1}{3}$, and the sum would be over two lines.

\figamba

The reasoning behind separating the first three and last Stokes parameters in Equations~\ref{eq:chisq} and \ref{eq:chisq2} comes from the strong-field limit of the Hanle effect.   
As already described, the 
first three Stokes parameters 
$IQU$ depend only on the direction
(e.g. unit vector $\uvec{b}$), and not
the magnitude $B$ of the magnetic field. On the other hand, Stokes $V$ parameters scale only with the magnitude $B$ of the magnetic field. Thus, the $\chi^2$ sorting needs to be separated into its two components, as shown in Equations~\ref{eq:chisq}-\ref{eq:chisq2}. 
We store in the database Stokes vectors $S_i$ computed only with $B=1$ G. The first 3 Stokes parameters, are determined by minimization of Equation~\ref{eq:chisq} yielding acceptable values of $\uvec{b}$, along with the smallest normed differences in integrated Stokes $V$, as given by a database search.
Once the direction $\hat{\bf b}$ is known, the contribution of Stokes $V$ to $\chi^2$ in Equation~\ref{eq:chisq2} is identically zero only when
\begin{eqnarray}
\nonumber
S_3&=&O_3, \mathrm{\ \ hence} \\
B &=& O_3/ S_3(B=1), \label{eq:algebraic}
\end{eqnarray}
which is the analytical solution for $B$ because $S_3 = B \cdot S_3$, where $B=1$.

Equation~\ref{eq:algebraic} then yields the magnetic field strength compatible with all the observed and computed Stokes parameters, without reference to the values $\sigma_3$ for each line.  
The value of $V$ used for estimating $B$ can be taken either from the strongest line or the weighted mean of a number of observed lines via a CLEDB configuration parameter.   
This procedure justifies argument number 4 listed above.

To sum up, the number of calculations needed becomes of the order of $10^6$ when using the discretization example in Table \ref{tab:discretization}, so that searches become fully tractable even on desktop computers.  Figures~\ref{fig:flowcledb} and \ref{fig:flow} show the overview and detailed CLEDB scheme as flowcharts.

\subsection{Performance}

In some initial tests using Python,  
solutions are obtained 
in 0.2 sec. for
 the parameters listed in Table~\ref{tab:discretization},
 using a fairly current off-the-shelf laptop like a 64-bit Macbook Pro with a 2.3 GHz Quad-Core Intel Core i7, with 16GB RAM. By compressing database files storage to 32 bit integers, 
we halve the disk space required, while 
incurring about 2 sec. of overhead each time the data are
read and decompressed. 
There is therefore a small advantage in finding all observations matching a given database value of $y_0$ before 
searching for solutions. CLEDB implements such a pre-search in its CLEDB\_PREPINV module, where for any measured cluster of $y$- heights, CLEDB searches and selects the nearest $y_0$ database position.

\figambb

Characteristics of the typical 
performance are shown in Figures~\ref{fig:amba} and \ref{fig:ambb}, applied to
the Fe XIII line pair at 1.0747 $\mu$m and 1.0798 $\mu$m. 
Figure~\ref{fig:amba} shows 
the derived physical parameters 
for a search of synthetic 
Stokes parameters drawn randomly from the database in the upper panel. The rms uncertainties 
assigned to the synthetic observations are for photon-counting noise associated with a total of 6 million counts 
in the brightest (1.0747 $\mu$m) line.  The lower panel shows the corresponding differences between the observed and computed Stokes 
parameters for these solutions.

Figure~\ref{fig:ambb} shows how the number of acceptable solutions varies with the noise levels.
As  
anticipated, sufficient counts must be accumulated to constrain 
the plasma and magnetic properties 
of coronal plasma using forbidden 
coronal lines.  Unanticipated is the result that $\approx 10^7$ 
counts are required to 
arrive at the minimally ambiguous set of solutions.  There is no benefit to accumulating more counts except that the magnitude of
$B$ can be better constrained using 
Equation~\ref{eq:algebraic}.  Also shown are estimates of the counts  
that might be accumulated  with 
a DKIST CRYO-NIRSP like
instrument in 1 second for a $0.5"\times0.5"$ region.  
Assuming that 
the instrument can achieve
photon-limited noise, 
a factor of 30 more
counts should be easily
achievable with longer integrations and spatial binning.  It remains to be seen what the nature of the noise of the instrument might be to affect the estimates given here. 
 \figtwos

 \section{Discussion}\label{S-discussion}
 
The CLEDB algorithm is centered on a straightforward 
least-squares  match of observed and computed $I,Q$, and $U$
Stokes parameters, which determine the magnetic field unit vector $\uvec{b}$. This is  
combined with 
magnetic field strength $B$ 
given algebraically by
the ratio of observed to computed $V$ parameters (Equation~\ref{eq:algebraic}).  
The algorithm uses 
line profiles 
integrated over 
frequency, which may include multiple components separated  perhaps  
 using multiple 
 Gaussian fits.
An
arbitrary number of two or more M1  coronal  
emission lines, each formed in the
strong-field limit of the Hanle effect \citep{Casini+Judge1999,Sahal}, can be used for a full vector solution, while a LOS magnetic approximation is available for one line.  
However, the physics dictates that the use of lines of the same ion  minimizes potentially damaging systematic errors.

The algorithm delivers the closest solutions to those in computed databases, including all solutions acceptable with the $\chi^2$- statistic 
for each measured component.  
Natural symmetries imply that 
at least two solutions are found for each component, even in the limit of negligible noise. To achieve 
this limit,  one test calculation (Figure~\ref{fig:amba}) required $>$ 6 million counts integrated along the line profile.  We 
also estimated that the CRYO-NIRSP instrument at the new DKIST observatory can achieve this with a combination of exposures as short as a few seconds, with modest spatial binning.

In Figure~\ref{fig:twos} we show the results of a numerical experiment in which we force the algorithm to return solutions from a situation from a scenario
entirely incompatible with a single source.  Two sources of equal intensity are placed, one at $x=-2$ in units of $R_\odot$, the other at ten points between 
$x=-2$ and 0. A total of $3\times 10^6$ counts were assumed to be accumulated in Stokes $I$.
To avoid confusion, we kept the 
magnetic field vector identical, seeking only to explore the 
ability of the algorithm to recognize through $\chi^2$ values that there is no single match in the database.  While this is a simple case, it is 
in one sense a ``worst case'' scenario in that the two sources are equally bright along the $x$- direction. As expected, the algorithm shows successes and failures. The solutions at $x=2$ have the smallest $\chi^2$, which increase almost monotonically with increasing source separation.  This is good news, as the algorithm not only recovers the correct solution when the sources are in the same location, but also $\chi^2$ increases significantly when the sources are separate.  Thus, the $\chi^2$ can show that there is indeed sufficient information in the spectra in order to reveal a poor fit. The other good news is the expected increase in mean electron density as the second source approaches $x=0$.  

However, the middle panel shows that the $x$- coordinates returned do not follow
anything approaching a linear trend. The line shows the position of a single source found at the same coordinates as the second source from above.
If the algorithm were linear in its response to the $x$- coordinate, 
then we would expect the points plotted to follow 
a line starting from (-2,-2) with half the slope shown.  Clearly, the algorithm is sufficiently non-linear to disallow the possibility of finding centers of emission along the LOS if two or more sources exist with the same or similar brightness. This is just as expected from the discussion in section 3.3 of \citet{Judge+Casini+Paraschiv2021}.  

We finish by making some general observations.
Our earlier work \citep{Judge+Casini+Paraschiv2021} clarifies how the present algorithm resolves earlier problems by
solving for the scattering geometry as well as the thermal
and magnetic parameters of the emitting plasma. The companion paper of this work \citep{par2022} will focus on benchmarking CLEDB on synthetic data, while waiting for the first full Stokes coronal observations to become available.

First, these inversions are far less dependent on the 
signal-to-noise ratios of the very weak Zeeman-induced Stokes $V$ profiles, a result contrasting with the earlier methods examined
\citep{Plowman_2014,Dima+2020}. While our solutions depend  linearly on the ratio of observed to computed $V$ values (Equation~\ref{eq:algebraic}),
the earlier solutions depend on the observed differences between measured $V$ values (see Equation~11 in \citealp{Dima+2020} and Equation~7 in
\citealp{Judge+Casini+Paraschiv2021}, which have correspondingly larger propagated uncertainties.
This is good news because the $V$ signals are small, being first order in the small parameter $\nu_L/\Delta\nu_D$. 
Secondly, it is clear that once applied, any user of this scheme is left to see which of the various solutions might make best sense when the pixel-to-pixel variations are taken into account, or if other constraints are available (e.g. independent knowledge of the geometry of the emitting plasma). 
This research area should be explored in the future, and may be ripe for application of machine-learning techniques.

Thirdly, using lines from the same ions in fact have advantages. We gain accuracy by using such ions without worrying about unknown factors such as temperatures, ionization fractions and abundances, and 
with this methodology we need not worry about the special degeneracies identified by \citet{Dima+2020}.  

Lastly, we note that because of the physical separation underlying Equations~\ref{eq:chisq}-\ref{eq:chisq2}, any independent knowledge of B$_{LOS}$ or $|B|$ can be easily
included in a CLEDB implementation. One example might be the use of oscillation data once the density is solved for from just IQU observations, in order to determine the value of $|B|$ from the observed oscillation phase speeds
\citep[see][]{2007Sci...317.1192T,2020ScChE..63.2357Y}.

\acknowledgements

The authors thank R. Casini for discussions and the careful reading and review of the initial submission. 
Furthermore, we are grateful for the anonymous reviewer's pertinent suggestions that improved this work.

\begin{dataavailability}
CLEDB and sample test data are available on Github via \\ \href{https://github.com/arparaschiv/solar-coronal-inversion}{github.com/arparaschiv/solar-coronal-inversion} or directly from the corresponding author on reasonable request. Furthermore, the CLEDB package provides detailed documentation. \href{https://github.com/arparaschiv/solar-coronal-inversion/blob/master/codedoc-latex/README-CODEDOC.pdf}{See README-CODEDOC.pdf}
\end{dataavailability}

\begin{fundinginformation}
A.R.P. was primarily funded for this work by the National Solar Observatory (NSO), a facility of the NSF, operated by the Association of Universities for Research in Astronomy (AURA), Inc., under Cooperative Support Agreement number AST-1400405. A.R.P. and P.G.J. are funded by the National Center for Atmospheric  Research, sponsored by the National Science Foundation under cooperative agreement No. 1852977. 
\end{fundinginformation}

\begin{conflict}
The authors declare that there is no conflict of interest.
\end{conflict}

\bibliographystyle{spr-mp-sola} 
\bibliography{ms.bib}
\end{article} 

\end{document}

%% file: defs.tex
\newcommand*{\rom}[1]{\expandafter\@slowromancap\romannumeral #1@}

\newcommand\ion[2]{#1$\;${%
\ifx\@currsize\normalsize\small \else
\ifx\@currsize\small\footnotesize \else
\ifx\@currsize\footnotesize\scriptsize \else
\ifx\@currsize\scriptsize\tiny \else
\ifx\@currsize\large\normalsize \else
\ifx\@currsize\Large\large
\fi\fi\fi\fi\fi\fi
\rom{#2}}\relax}%

\newcommand\pmag{\ifmmode \omega_{\rm L}/\Delta\omega_{\rm D}
\else$\omega_{\rm L}/\Delta\omega_{\rm D}$\fi}
\renewcommand{\v}[1]{\ifmmode {1\over2}(3\cos^2#1-1)\else ${1\over2}(3\cos^2#1-1)$\fi}
\renewcommand{\v}[1]{\ifmmode {v}(#1)\else $ {v}(#1)$\fi}
\newcommand{\tb}{\ifmmode \Theta_{\rm B}\else $\Theta_{\rm B}$\fi}
\newcommand{\gb}{\ifmmode \gamma_{\rm B}\else $\gamma_{\rm B}$\fi}
\newcommand{\vtb}{\ifmmode \vartheta_{\rm B}\else $\vartheta_{\rm B}$\fi}
\newcommand{\vpb}{\ifmmode \varphi_{\rm B}\else $\varphi_{\rm B}$\fi}
\newcommand{\vtm}{\ifmmode \vartheta_{\rm M}\else $\vartheta_{\rm M}$\fi}
\newcommand{\pop}[1]{\ifmmode N(  {\scriptstyle #1}) \else $N(  {\scriptstyle #1})$\fi}
\newcommand\ft{l(\vtm)}
\newcommand\alig{k_J(T_e,n_e,\vtm)\,\v{\vtb} \,}
\newcommand\modalig{k_J(T_e,n_e,\vtm)|\v{\vtb}| \,}

\newcommand\freq{\nu}

\newcommand{\pgj}[1]{\textbf{ \textcolor{red}{#1}}}

\def\UrlFont{\sf}            


\newcommand{\BibTeX}{\textsc{Bib}\TeX}
\newcommand{\etal}{{\it et al.}}

\renewcommand{\vec}[1]{{\mathbfit #1}}
\newcommand{\deriv}[2]{\frac{{\mathrm d} #1}{{\mathrm d} #2}}
\newcommand{\rmd}{ {\ \mathrm d} }
\newcommand{\uvec}[1]{ \hat{\mathbf #1} }
\newcommand{\pder}[2]{ \f{\partial #1}{\partial #2} }
\newcommand{\grad}{ {\bf \nabla } }
\newcommand{\curl}{ {\bf \nabla} \times}
\newcommand{\vol}{ {\mathcal V} }
\newcommand{\bndry}{ {\mathcal S} }
\newcommand{\dv}{~{\mathrm d}^3 x}
\newcommand{\da}{~{\mathrm d}^2 x}
\newcommand{\dl}{~{\mathrm d} l}
\newcommand{\dt}{~{\mathrm d}t}
\newcommand{\intv}{\int_{\vol}^{}}
\newcommand{\inta}{\int_{\bndry}^{}}
\newcommand{\avec}{ \vec A}
\newcommand{\ap}{ \vec A_p}

\newcommand{\bb}{\vec B}
\newcommand{\jj}{ \vec j}
\newcommand{\rr}{ \vec r}
\newcommand{\xx}{ \vec x}

\newcommand{\adv}{    {\it Adv. Space Res.}} 
\newcommand{\annG}{   {\it Ann. Geophys.}} 
\newcommand{\aap}{    {\it Astron. Astrophys.}}
\newcommand{\aaps}{   {\it Astron. Astrophys. Suppl.}}
\newcommand{\aapr}{   {\it Astron. Astrophys. Rev.}}
\newcommand{\ag}{     {\it Ann. Geophys.}}
\newcommand{\aj}{     {\it Astron. J.}} 
\newcommand{\apj}{    {\it Astrophys. J.}}
\newcommand{\apjl}{   {\it Astrophys. J. Lett.}}
\newcommand{\apss}{   {\it Astrophys. Space Sci.}} 
\newcommand{\cjaa}{   {\it Chin. J. Astron. Astrophys.}} 
\newcommand{\gafd}{   {\it Geophys. Astrophys. Fluid Dyn.}}
\newcommand{\grl}{    {\it Geophys. Res. Lett.}}
\newcommand{\ijga}{   {\it Int. J. Geomagn. Aeron.}}
\newcommand{\jastp}{  {\it J. Atmos. Solar-Terr. Phys.}} 
\newcommand{\jgr}{    {\it J. Geophys. Res.}}
\newcommand{\mnras}{  {\it Mon. Not. Roy. Astron. Soc.}}
\newcommand{\nat}{    {\it Nature}}
\newcommand{\pasp}{   {\it Pub. Astron. Soc. Pac.}}
\newcommand{\pasj}{   {\it Pub. Astron. Soc. Japan}}
\newcommand{\pre}{    {\it Phys. Rev. E}}
\newcommand{\solphys}{{\it Solar Phys.}}
\newcommand{\sovast}{ {\it Soviet  Astron.}} 
\newcommand{\ssr}{    {\it Space Sci. Rev.}} 
\chardef\us=`\_

\newcommand{\figsym}{
\begin{figure}[ht]
\includegraphics[width=1.0\linewidth]{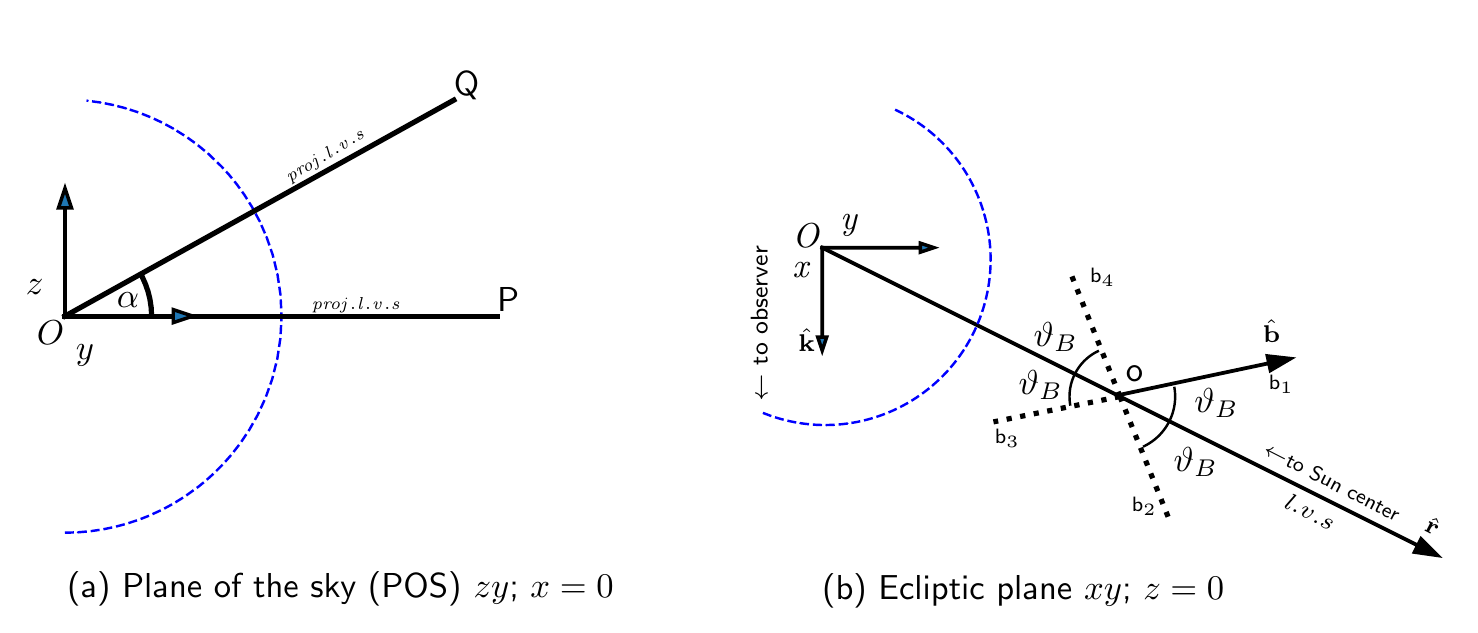}
\caption{ 
Plane of sky and ecliptic plane geometry for coronal emitting structures. The \textit{blue dashed curve} is the solar surface.  The $y$- coordinate of the emitting plasma denotes the elongation.
\textit{Panel (a)} shows 
the geometry in the POS  ($x=0$).  Points $P$ and 
$Q$ see identical photospheric radiation
from the Sun centered at $O$. 
The reference direction for linear polarization in the local vertical of the Sun [\textit{l.v.s.}] is assumed throughout to be along the $z-$ axis. 
\textit{Panel (b)} shows how points $b_1,b_2,b_3$ and $b_4$ along the vectors 
{$\protect\overrightarrow{ob_1},\protect\overrightarrow{ob_2},\protect\overrightarrow{ob_3},\protect\overrightarrow{ob_4}$} experience the same 
value of $\cos^2 \vartheta_B$, the angle between the local magnetic vector in the corona and 
 the $l.v.s.$\textit{; the line joining the point to Sun center. }
At the angles traced along these \textit{dotted lines}, and for
the cones rotated about the $l.v.s.$, solutions of the statistical equilibrium equations are the same as for the \textit{solid line}, that is when the saturated Hanle regime assumption holds. 
}
\label{fig:sym}
\end{figure}
}

\newcommand{\figspheresm}{
\begin{figure}[ht]
\includegraphics[width=1.\linewidth]{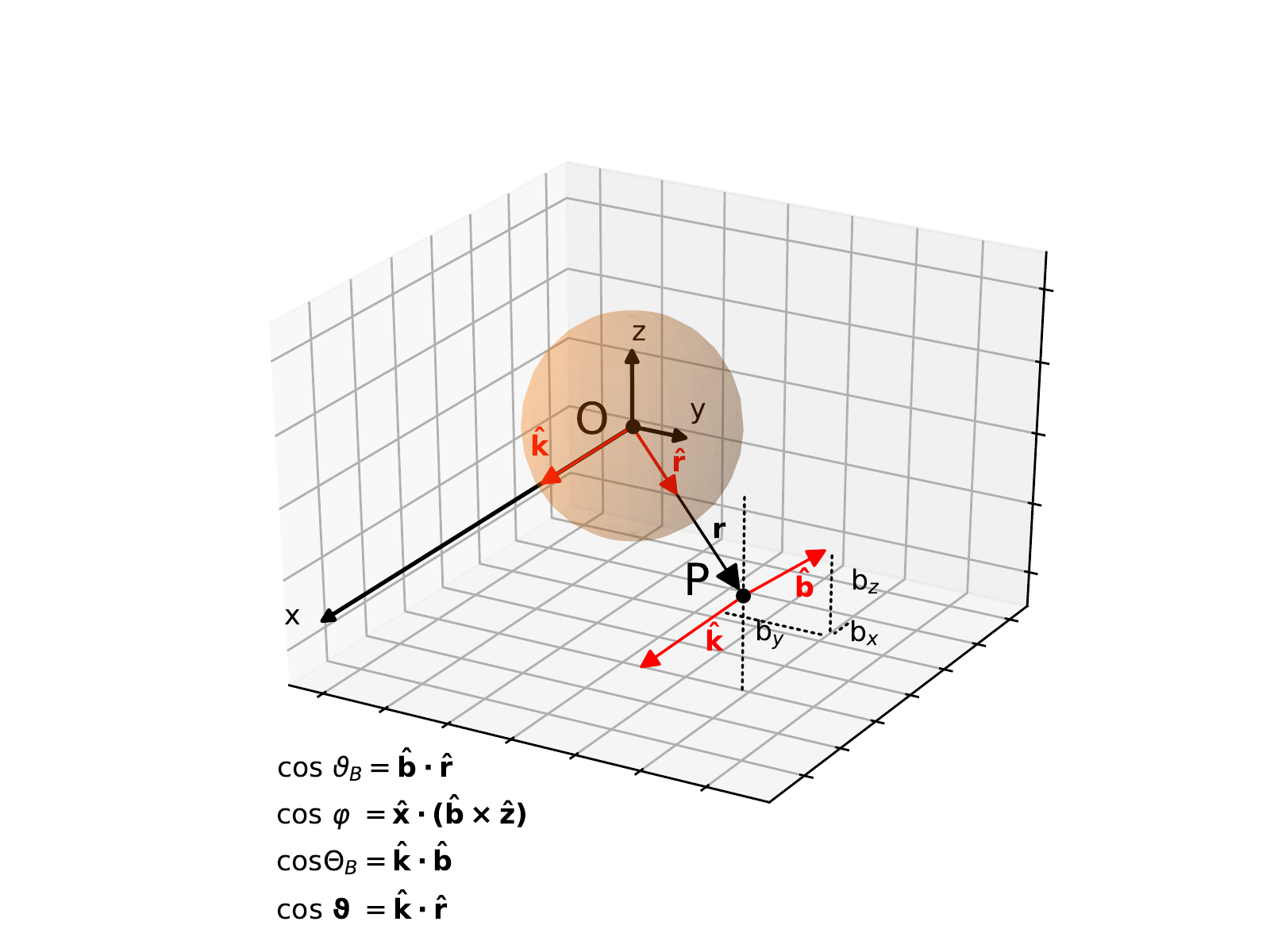}
\caption{The scattering geometry of point $P$ is shown in the observer's frame with projections of the magnetic field components in the same frame.  
The LOS through Sun center lies along unit vector $\mathbf{\hat k}$, which is parallel to the $x$-axis.  
We assume that the LOS through the corona are parallel to $\mathbf{\hat k}$.  
$\Phi_B$ ($=\pi-\gamma_B$ in CJ99)  is defined by $\arctan \hat b_z / \hat b_y$ if we take the reference direction for linear polarization along the  $z$- axis. The unit vectors of interest are marked with \textit{red arrows.}
}
\label{fig:spheresm}
\end{figure}
}

\newcommand{\figflow}{
\begin{figure}[t]
\includegraphics[trim=0cm 9cm 0cm 1cm, clip, width=1.\linewidth]{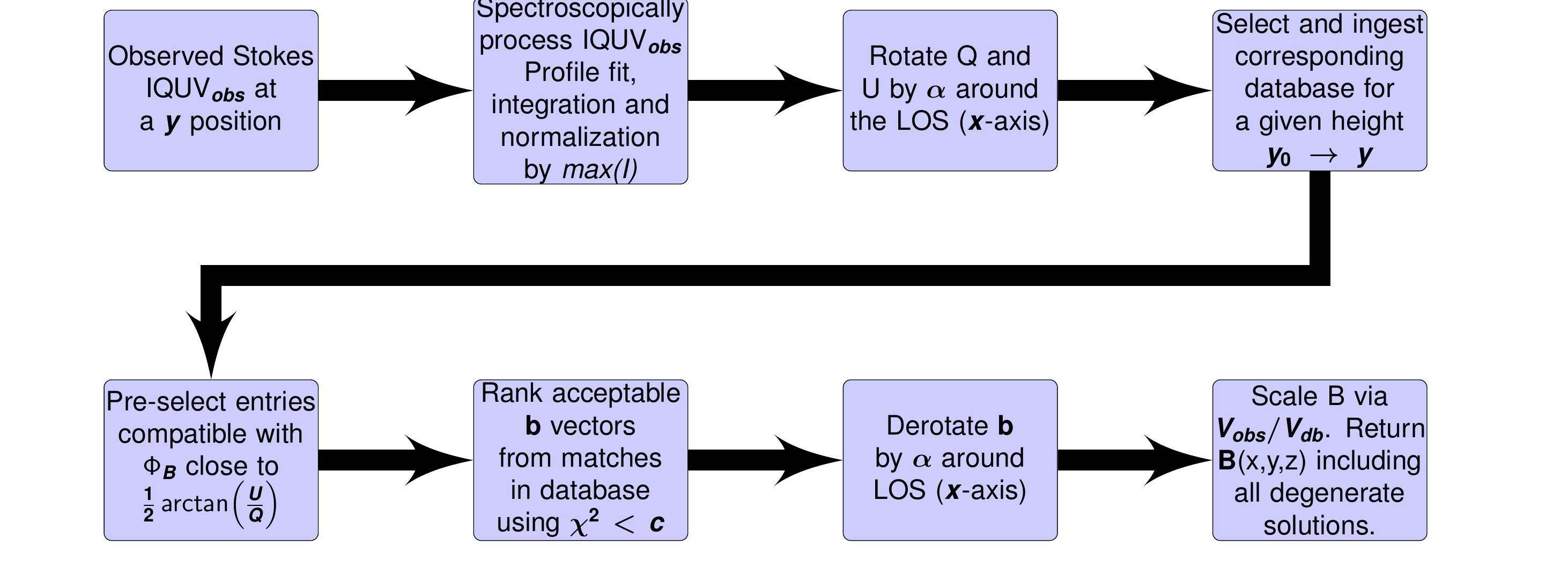}
\caption{CLEDB 2-line magnetic inversion algorithm flowchart that is a subcomponent of the CLEDB\_PROC module.  An important aspect is the delivery of multiple 
possible solutions for each observation at the last step.
Note that the $x$- coordinate of the point in space,  as well
as nearest electron density, are returned along with $\mathbf{B}$.  The figure 
uses the notation $V_{\mathrm{obs}}$ and $V_{\mathrm{db}}$ for observed and computed values of the amplitudes of the Stokes parameters corresponding to $O_3$ and $S_3 (B=1)$ in the text.
}
\label{fig:flow}
\end{figure}
}

\newcommand\tabangles{
\protect\begin{table}
\begin{tabular}{lrll}
\hline
Angle & definition & source & notes\\
\hline
    $\alpha$ & arctan$(z/y)$ & this work & Figure 1(a)\\
    $\phi$ &  $\arccos
    \left( \mathbf{\hat x \cdot (
    \hat b \times \hat z)} \right) $ & this work& azimuthal angle of $\mathbf B$\\
    $\theta$ & arccos$(\mathbf{ \hat z \cdot \hat b})$ & this work&polar angle of $\mathbf B$  \\
    \boldmath$\vartheta$& arccos$(\uvec{k}\cdot\uvec{r})$    &CJ99 fig. 5 & (angle between LOS and  l.v.s.) \\
    \boldmath$\vartheta_B$& $\arccos(\mathbf{ \hat r \cdot \hat b})$    &CJ99 fig. 5 & (See also Figure 1(b))   \\
    \boldmath$\varphi_B$ & $\arccos\left (\mathbf{ (\hat k \times \hat r) \cdot (\hat b \times \hat r)} \right)$    &CJ99 fig. 5 & \\
    \boldmath$\Theta_B$& arccos$(\mathbf{ \hat k \cdot \hat b})^\ast$ & CJ99 fig. 5 &  \\
      \boldmath$\Phi_B$& $\frac{1}{2}\arctan(U/Q)$ & CJ99 fig. 5 &
\\
\hline    
\end{tabular}
    \caption{Angle definitions
    in the l.v.s frame except where noted}
$^\ast$As observed from Earth, the LOS
and $x$- axis are almost 
parallel. For example an elongation 
$\sqrt{y^2 +z^2}$ of $2R_\odot$ the angle between the LOS and $\uvec{k}$ is 
only has 
$0.0093 $ radians 
(0.5$^o$).   Any errors introduced with this assumption are minor compared with the other observational and theoretical challenges 
presented by the problem at hand.
    \protect\label{tab:angles}
\end{table}
}

\newcommand\tabsols{
\begin{table}
\caption{2-line CLEDB solution output products compared with synthetic observations.} The CLEDB solutions are degenerate to 180$^\circ$ with respect to the LOS angle components. The CLE atmosphere parameters provide the ground-truth reference values. These are averaged along the LOS. H and D refer, respectively, to the coronal height above the limb and the LOS depth in units of R$_\odot$. The electron density is given in $\log 10$ cm$^{-3}$, magnetic angles in degrees, and magnetic field components in G.
\label{tab:table1}
\begin{tabular}{ccccccccccc}
\hline
Index & $\chi^2$ & n$_e$ & H & D & $|$B$|$ & $\Phi_B$ & $\Theta_B$ &  B$_x$ & B$_y$ & B$_z$ \\
5982696 & 2.12e-2  & 8.10 & 1.18 & -0.52 &  6.0 & 108.3 &  72.8 & -1.85 &  5.47 &  1.72 \\
6669594 & 2.12e-2  & 8.05 & 1.18 &  0.55 &  6.0 & 253.3 & 108.9 & -1.63 & -5.47 & -1.93 \\
5990814 & 2.13e-2  & 8.10 & 1.18 & -0.52 & -6.0 & 289.4 & 108.9 & -1.93 &  5.47 &  1.93 \\
6661476 & 2.13e-2  & 8.05 & 1.18 &  0.55 & -6.0 &  72.2 &  72.8 & -1.71 & -5.52 & -1.72 \\
5982606 & 3.65e-2  & 8.10 & 1.18 & -0.52 &  6.8 & 106.6 &  72.8 & -1.88 &  6.28 &  2.02 \\
6669684 & 3.65e-2  & 8.05 & 1.18 &  0.55 &  6.8 & 255.0 & 108.9 & -1.68 & -6.28 & -2.29 \\
5982697 & 5.64e-2  & 8.10 & 1.18 & -0.52 &  5.9 & 108.3 &  74.5 & -1.81 &  5.44 &  1.57 \\
6669593 & 5.64e-2  & 8.05 & 1.18 &  0.55 &  5.9 & 253.3 & 107.1 & -1.61 & -5.44 & -1.74 \\
5990904 & 5.79e-2  & 8.10 & 1.18 & -0.52 & -5.4 & 291.2 & 108.9 & -1.94 &  4.89 &  1.84 \\
6661386 & 5.79e-2  & 8.05 & 1.18 &  0.55 & -5.4 &  69.9 &  72.8 & -1.75 & -4.91 & -1.60 \\
\hline
CLE : &        & 7.97 & 1.18 & -0.49 &  6.3 &       &       & -1.59 &5.72 & 2.02 \\
\hline
\end{tabular}
\end{table} 
}

\newcommand{\tabdisc}{
\begin{table}
    \begin{center}
\begin{tabular}{llr}
\hline
quantity & number & range \\
\hline
 $\log_{10}n_e/n_0^\dag(r)$ (electron density) &$n_{n_e}=15$ &$-2\rightarrow +2
   $ \\
    $x$- axis (LOS, units $R_\odot$)     & $n_x=100$ &
    $-2.5 \to 2.5$\\
    $\phi$   & $n_\phi$ = 180 & $0\to 2\pi$ \\    
    $\theta$ & $n_\theta$ = 90 & $0\to \pi$\\
    $\bar\theta$ & $n_{\bar\theta}\approx 10$ & $0\to \pi$\\
    $N=  n_{n_e}n_x  n_\phi n_\theta $
     & 24.3 million&\\
        $N'=  n_{n_e}n_x  n_\phi n_{\bar\theta}$
     & 2.7 million&\\
     \hline
     Size of each database file of 2 lines, each with  & 388 MB \\
     4 Stokes parameters stored as 32 bit integers\\
\hline
\end{tabular}
            \end{center}
   \caption{Example of numerical discretization
    for each database file at elongation  $y_0$.}
\vspace{-0.4cm}The $N$ and $N'$ represent the sizes of databases as computed with either $n_{\theta}$ or $n_{\bar\theta}$. These correspond respectively to all database orientations and to the subset compatible with equation~\ref{eq:qu}. Each of the two discretization options can be user-selected via CLEDB control parameters.
\vspace{-0.9cm}
        \label{tab:discretization}
\end{table}
}

\newcommand{\figamba}{
\begin{figure}[ht]
\includegraphics[width=0.7\linewidth]{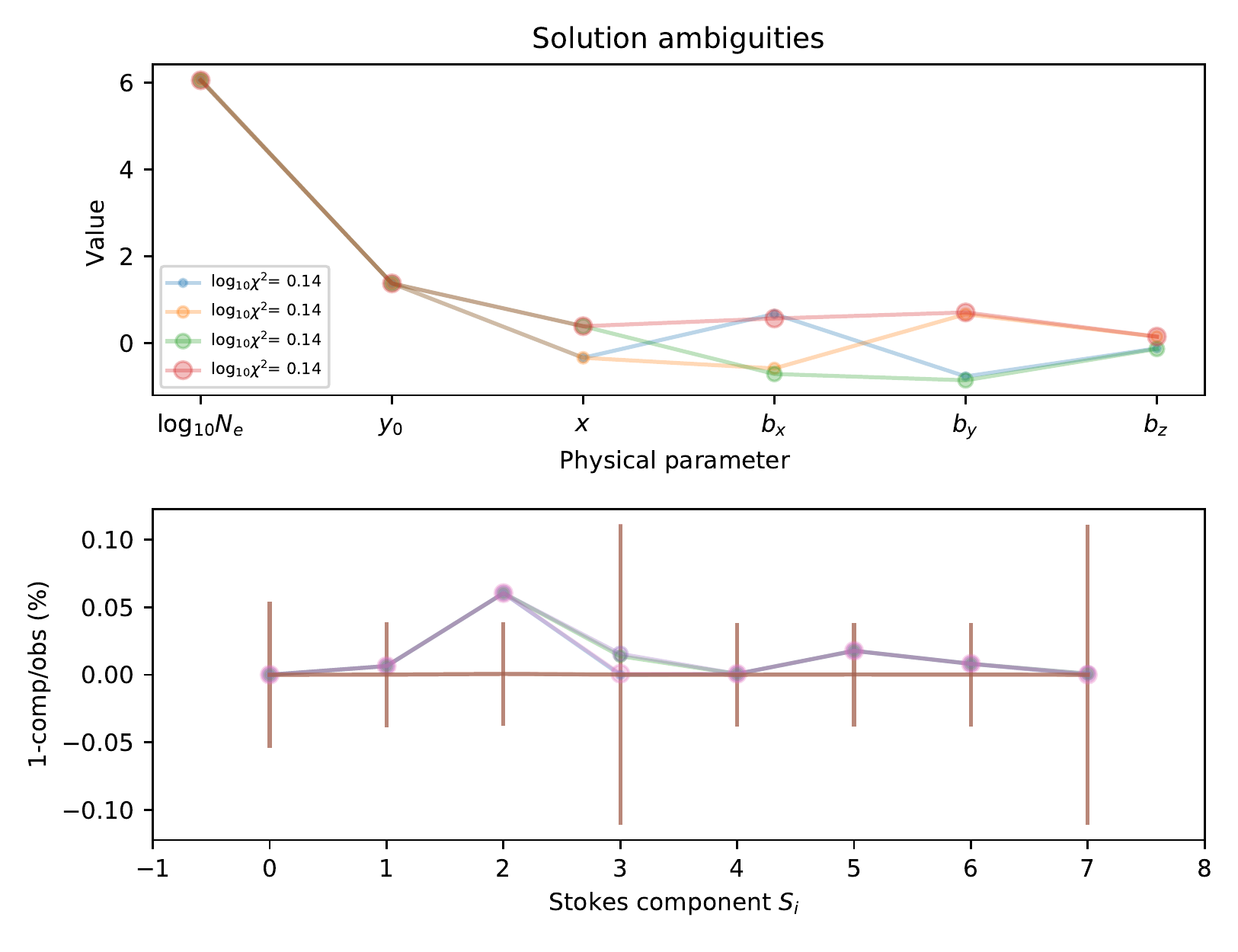}
\caption{ The figure shows four solutions computed by the database search algorithm that are compatible ($\chi^2 < 1.4$) with synthetic Stokes 
components of both the 
Fe XIIII 1.0474 $\mu$m and 1.0798 $\mu$m 
lines.  The synthetic data are for a randomly chosen model calculation with statistical errors
from 63 million counts in the 1.0747 $\mu$m line. The l\textit{ower panel} shows the differences in percent 
of the Stokes parameters in the observations, associated error bars, and solutions found from the search algorithm.}
\label{fig:amba}
\end{figure}
}

\newcommand{\figambb}{
\begin{figure}[ht]
\centering
\includegraphics[width=0.7\linewidth]{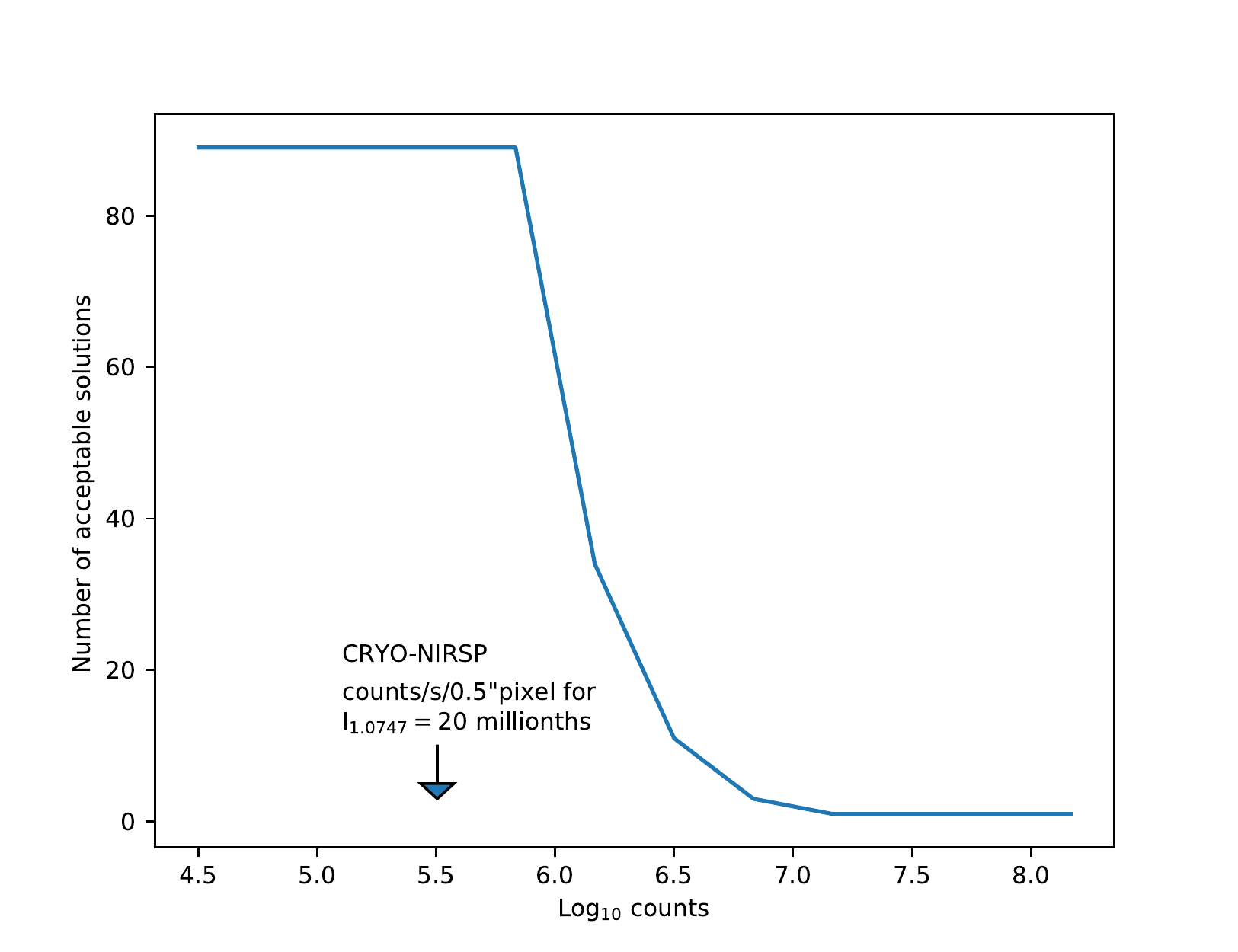}
\caption{ The  number of solutions $\chi^2 < 1.4$ compatible with 
a randomly chosen database entry as a function of the accumulated counts 
over the entire Fe XIII 1.0747 $\mu$m line is shown.  Also marked is an estimate of the counts accumulated 
by a CRYO-NIRSP like instrument on DKIST in 1 second over a $0.5"\times 0.5"$ region.  
A brightness of 20 millionths of the solar disk brightness, over the line profile was assumed as typical of the lower corona following \citet{Judge1998}. For a 10 second 
integration time, CLEDB would yield 
just four solutions compatible with such data, when considering the above defined $\chi^2$ limit. 
}
\label{fig:ambb}
\end{figure}
}
 
\newcommand{\figlosa}{
\begin{figure}[ht]
\tiny
\includegraphics[width=0.7\linewidth]{fig6_ecliptic.png}\vspace{-1.0cm}
\includegraphics[width=0.95\linewidth]{fig6_ATMOS_out.png}

\caption{Example of multiple superposed LOS magnetic dipole structures. The structures apparently resemble a simplified limb loop structure commonly seen above AR. The LOS integrated physical information is presented. Upper panel shows the three loops as seen along the LOS from the Ecliptic plane.  Three structures A,B and C are generated at different LOS positions. Structure A represents a double loop system. The bottom panel shows the individual physical parameters of each of the three dipolar structures. It is observed that these structures are not generated at equal heights along the vertical (z) axis. An apparent current sheet structure appears near the equatorial region. \arp{ARP: Everything will be rehauled, labels added and everything if the physics and the idea is what you wanted.}}
\label{fig:los1}
\end{figure}
}

 \newcommand{\figlosb}{
\begin{figure}[ht]
\tiny
\includegraphics[width=0.7\linewidth]{fig6_StokesIVQU.png}\vspace{-1.2cm}
\includegraphics[width=0.95\linewidth]{fig6_line_spectra.png}

\caption{Example of multiple superposed LOS magnetic dipole structures. The full Stokes Fe-XIII 1.074$\mu$ spectral information shown here is corresponding to the geometry and physical parameters described in fig. \ref{fig:los1}.  The upper panel shows synthesized integrated Stokes $IQUV$ emission, and the region for which detailed spectra are plotted. The bottom panel depicts the spectral information corresponding to a representative pixel (y,z) = (1.04,0.01)$R_{\odot}$ along an apparent equatorial current sheet (CS). In this particular location, all three independent magnetic structures A, B, and C contribute to the observed LOS integrated parameters. \arp{ARP: Everything will be rehauled, labels added and everything if the physics and the idea is what you wanted.}}
\label{fig:los2}
\end{figure}
}

\newcommand{\figtwos}{
\begin{figure}[ht]
\centering
\includegraphics[width=.6\linewidth]{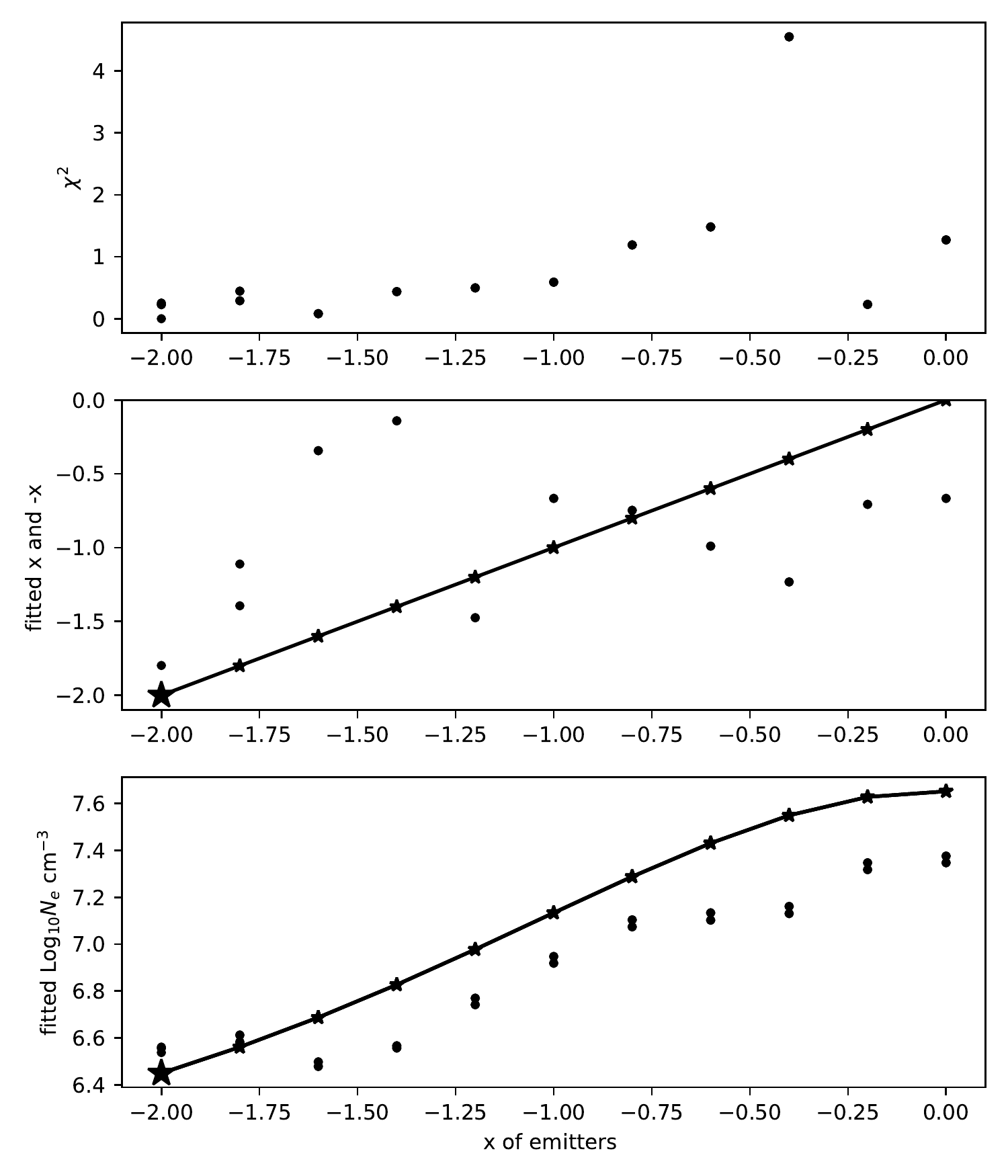}
\caption{
The figure shows values of $\chi^2$ (\textit{top}), the fitted value of LOS coordinate $x$ (\textit{middle}) and the electron density (\textit{bottom}) matched to emergent spectra of \textit{two} sources of emission in the two Fe~XIII M1 emission lines.  These two sources have different positions along the 
LOS, and densities.  Thus they 
violate the basic single-source assumption of the algorithm. This illustrates the kind of errors made when 
two sources contribute to the emission from one pixel. 
One source is held fixed at 
 $x=-2$ (\textit{large star symbol}) and the other placed at ten different positions with $x$- coordinates between -2.0 and 0.0 $R_\odot$
(\textit{smaller star symbols joined by the solid black line}, plotted both for $x$ and electron density as simple scalar examples). 
The  $\chi^2$ values increase almost monotonically with
the $x$- coordinate of the second source.   The returned $x$- coordinates are different from the mean source position (between $x=-2$ and the 
\textit{small starred line}).
As the second point is moved from -2.0 to 0.0 in $x$, the algorithm returns a density increasing almost monotonically, following a weighted
average of the densities of the two sources. }
\label{fig:twos}
\end{figure}
}

\newcommand{\figcledb}{
\begin{figure}[h]
\includegraphics[width=1\linewidth]{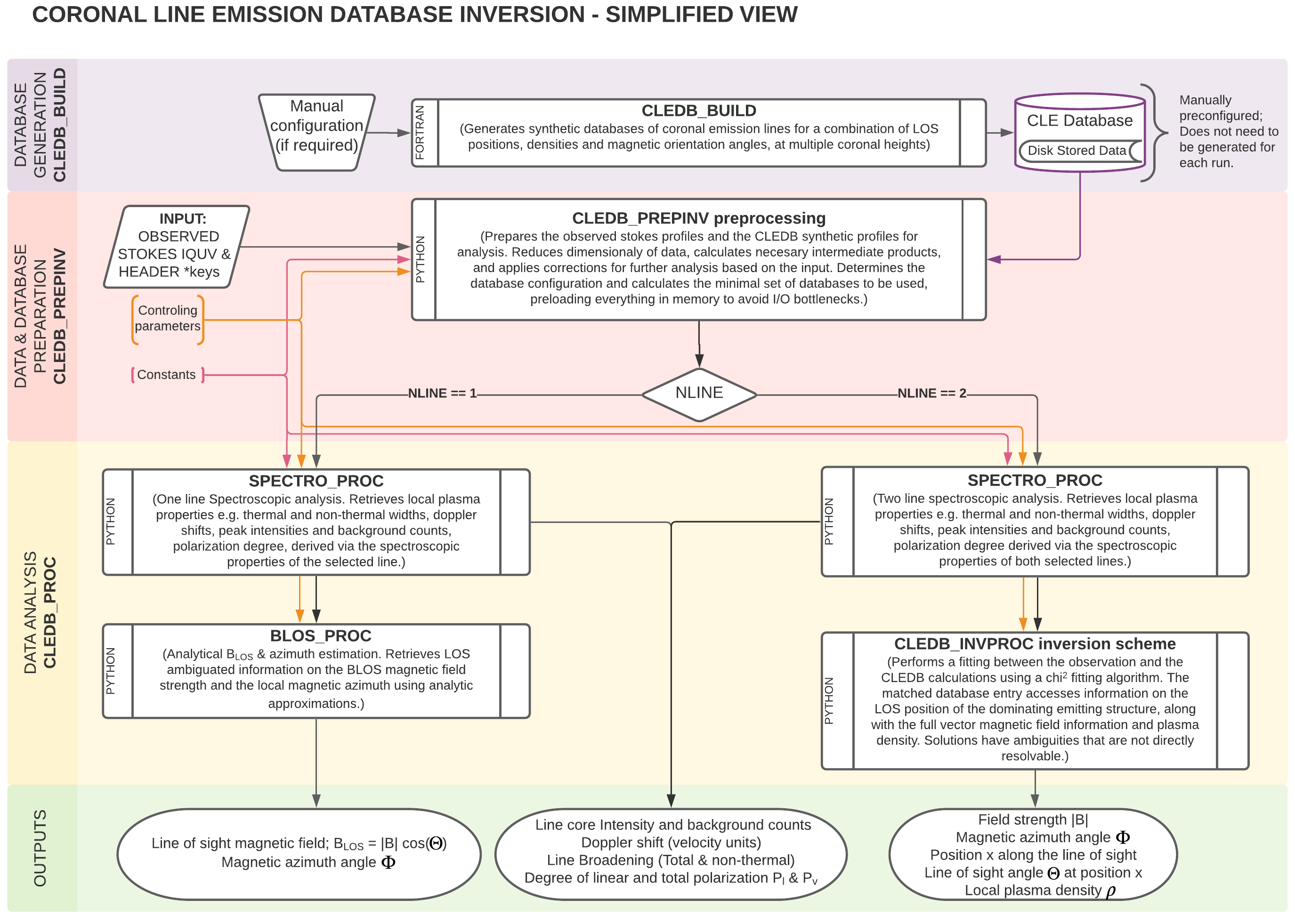}
\caption{A simplified view of the CLEDB module scheme. Three modules each focus on: the database generation via CLEDB\_BUILD; the CLEDB\_PREPINV data pre-processing; and the  CLEDB\_PROC data analysis and inversion.  The CLEDB\_BUILD module requires the execution of precompiled Fortran binaries of the CLE code due to computation speed requirements. Ideally, databases are only generated once per system based on user requirements. The other two modules are based on Python. The data analysis is split into two branches with different output products based on wether 1-line or 2-line Stokes IQUV data is ingested. An in-depth module description can be found in the documentation linked in the data availability section.}
\label{fig:flowcledb}
\end{figure}
}